\documentclass[sigconf]{acmart}

\usepackage{enumitem}
\usepackage[ruled, linesnumbered]{algorithm2e}
\usepackage{setspace}
\usepackage{graphicx}
\usepackage{multirow}

\AtBeginDocument{%
  }

\copyrightyear{2027}
\acmYear{2027}
\setcopyright{cc}
\setcctype{by}
\acmConference[ASPDAC '27]{32nd Asia and South Pacific Design Automation Conference}{January 25--28, 2027}{Tokyo, Japan}
\acmBooktitle{32nd Asia and South Pacific Design Automation Conference (ASPDAC '27), January 25--28, 2027, Tokyo, Japan}
\acmDOI{10.1145/3840404.3848215}
\acmISBN{979-8-4007-2923-2/2027/01}

\begin{document}

\title{Phase-Aware Monte Carlo Tree Search with LLM-Guided Proposal for Multi-Objective Physical Design Tuning}

\author{Jongho Yoon}
\orcid{0000-0002-9295-4850}
\affiliation{
  \department{Institute of AI}
  \institution{Pohang University of Science and Technology}
  \city{Pohang}
  \country{Republic of Korea}}
\email{jh.yoon@postech.ac.kr}

\author{Jeonghyun Choi}
\orcid{0009-0004-6261-7986}
\affiliation{
  \department{Department of EE}
  \institution{Pohang University of Science and Technology}
  \city{Pohang}
  \country{Republic of Korea}}
\email{choidh2013@postech.ac.kr}

\author{Jaemin Seo}
\orcid{0009-0000-1561-9170}
\affiliation{
  \department{Department of EE}
  \institution{Pohang University of Science and Technology}
  \city{Pohang}
  \country{Republic of Korea}}
\email{seojm@postech.ac.kr}

\author{Seokhyeong Kang}
\orcid{0000-0003-3015-1806}
\affiliation{
  \department{Department of EE}
  \institution{Pohang University of Science and Technology}
  \city{Pohang}
  \country{Republic of Korea}}
\email{shkang@postech.ac.kr}

\renewcommand{\shortauthors}{Yoon et al.}

\begin{abstract}
Physical design tuning must search a large parameter space under a limited budget of expensive full-flow executions.
The task is hard because design decisions are phase-coupled, failures may surface only in later stages, and multiple PPA objectives must be traded off over clean-feasible solutions.
We present \textbf{PILOT-Agent}, a phase-aware framework that formulates this problem as Monte Carlo Tree Search over configuration prefixes aligned with the ordered stages of the design flow.
At each prefix, an LLM Proposal Agent generates context-conditioned next-phase candidates.
A surrogate screening policy predicts which candidates are worth executing, and an archive-aware tree policy allocates the evaluation budget to candidates expected to improve the Pareto front.
Under the same budget, PILOT-Agent recovers clean-feasible Pareto fronts more accurately than Bayesian optimization and LLM-agent baselines, cutting hypervolume and coverage error by 31\% and 52\% over the strongest baselines.
PILOT-Agent also consistently finds design points improving one PPA objective without degrading the others.
\end{abstract}




\maketitle

\section{Introduction} \label{sec1}

Modern physical design flows expose many tunable parameters that critically affect power, performance, and area (PPA), and often determine whether the flow completes cleanly~\cite{jung2021metrics}.
As design complexity grows, manual tuning becomes impractical, motivating design space optimization (DSO) methods that search for high-quality configurations under limited budgets of expensive flow executions~\cite{geng2023ptpt,zheng2023boosting}.
Physical design tuning poses three coupled challenges.
First, the flow is sequential, so later outcomes depend on earlier phase decisions~\cite{ajayi2019openroad,agnesina2023parameter}.
Second, each evaluation is expensive, and weak configurations may fail only late in the flow, after substantial runtime has already been spent.
Third, designers seek not a single best configuration but a set of clean-feasible solutions, i.e., flow-complete with zero design rule violations, that covers competing PPA trade-offs~\cite{geng2023ptpt,zheng2023boosting,xu2024ranktuner}.

Prior work on physical design tuning has largely followed traditional optimization paradigms, including heuristic search~\cite{esbensen1996design,palesi2002multi}, Bayesian optimization (BO)~\cite{geng2023ptpt,luo2024attention,zheng2023boosting,jung2021metrics,Agnesina2023autodmp}, and reinforcement learning (RL)~\cite{agnesina2023parameter}.
However, many heuristic and BO methods search the joint parameter space without explicit phase modeling, making late failures difficult to trace to earlier decisions.
RL naturally supports sequential interaction, but prior formulations target a single stage rather than tuning across the full flow under a limited evaluation budget.
Because the phase structure determines where failures emerge and which partial decisions deserve further budget, these challenges must be handled jointly within a single search loop.

Recent large language model (LLM)-based approaches guide expensive design optimization as iterative tuning agents or retrieval-augmented advisors~\cite{ghose2025orfsagent,pan2025crop}, or couple LLM guidance with an existing optimizer such as BO~\cite{yin2024adollm}.
These methods show that LLMs can inject useful design priors and flexibly propose configurations.
Even so, two issues remain unresolved: organizing proposal and credit assignment around the phase structure of the flow, and identifying which proposals merit expensive full-flow evaluation.

We propose \textbf{PILOT-Agent}\footnote{Source code: \url{https://github.com/jhy00n/PILOT-Agent}}, a phase-aware framework that handles all three challenges in one search loop.
PILOT-Agent explores configuration prefixes, i.e., parameter assignments for phases decided so far, using \emph{Monte Carlo Tree Search} (MCTS).
This tree structure captures phase coupling and lets complete runs update earlier decisions.
At each prefix, a pretrained LLM, used zero-shot as a \emph{Proposal Agent}, broadens the candidate set with next-phase proposals.
A lightweight \emph{surrogate screening policy} and an archive-aware tree policy then decide which candidates justify an expensive full-flow run.
Our contributions are as follows:
\begin{itemize}[leftmargin=0.5cm]
    \item We formulate physical design tuning as \textbf{phase-prefix search} and develop a \textbf{phase-aware MCTS framework} that branches on next-phase decisions while scoring each branch by its completed configuration.
    \item We introduce \textbf{context-conditioned LLM proposals}: an LLM Proposal Agent broadens the local candidate set at each prefix by conditioning on the objective, the current prefix, and measured search-state feedback.
    \item We design \textbf{surrogate-guided selection}: a surrogate screening policy scores completed rollouts by predicted clean-feasibility and expected hypervolume gain before the archive-aware tree policy allocates expensive evaluations.
    \item PILOT-Agent reduces average HV-Error by 31\% and ADRS by 52\% relative to the strongest baseline on each metric, while achieving the highest average feasible rate.
\end{itemize}
\section{Preliminaries} \label{sec2}

\subsection{Problem Formulation and Objective}
\label{subsec:problem_formulation}

We model physical design tuning as a phase-structured optimization problem over the ordered stages of the design flow, where a phase corresponds to a major flow stage such as synthesis, floorplanning, placement, clock-tree synthesis, or routing. 
Let the full configuration be:
\begin{equation}
\label{eq:full-configuration}
\mathbf{x}=\big(\mathbf{x}^{(1)},\mathbf{x}^{(2)},\dots,\mathbf{x}^{(P)}\big), \quad \mathbf{x}^{(p)}\in\mathcal{X}^{(p)},
\end{equation}
where \(\mathbf{x}^{(p)}\) denotes the parameter assignment for phase \(p\), \(\mathcal{X}^{(p)}\) is the corresponding search space, and \(P\) is the number of optimization phases. 
Evaluating a complete configuration \(\mathbf{x}\) returns a run status. 
When the flow completes, it also returns the PPA objective vector \(\mathbf{f}(\mathbf{x})=(\mathrm{Power},\mathrm{ECP},\mathrm{Area})\) and the final design rule violation (DRV) count.
Here, the effective clock period (ECP) is the target clock period ($T_{\mathrm{clk}}$) degraded by the worst negative slack (WNS), \(\mathrm{ECP}=T_{\mathrm{clk}}-\mathrm{WNS}\).
A run is \emph{clean-feasible} if the full flow terminates normally and produces a valid final layout with zero DRVs.

\begin{figure}[t]
\begin{center}
\vspace{-0.2cm} 
\includegraphics[width=80mm]{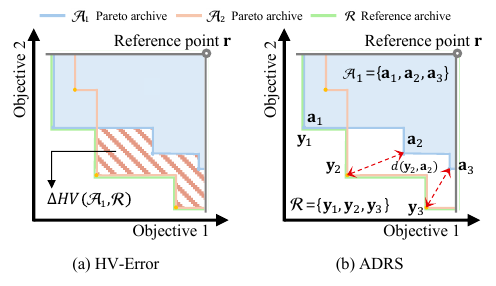}
\vspace{-0.4cm}
\caption{(a)~HV-Error: the shaded region shows the hypervolume gap $\Delta HV(\mathcal{A}_1,\mathcal{R})$ between the measured archive~$\mathcal{A}_1$ and reference archive~$\mathcal{R}$. (b)~ADRS: the average nearest distance from each point in~$\mathcal{R}$ to~$\mathcal{A}_1$.}
\label{fig:preliminaries}
\end{center}
\vspace{-0.1cm}
\end{figure}

Given a budget of at most \(B\) full-flow evaluations, our objective is a high-quality clean-feasible Pareto archive over the PPA objectives.
Fig.~\ref{fig:preliminaries} illustrates the two archive-quality measures we use, hypervolume error (HV-Error) and average distance to reference set (ADRS), which capture dominated-volume loss and reference-front coverage, respectively.
Let \(\mathcal{A}\) denote the nondominated objective vectors \(\mathbf{f}(\mathbf{x})\) obtained from clean-feasible runs under the budget.
The hypervolume of \(\mathcal{A}\) under a reference point \(\mathbf{r}\), chosen componentwise worse than all scored points, is:
\begin{equation}
\label{eq:hv}
HV(\mathcal{A};\mathbf{r})=
\operatorname{Vol}\Big(\bigcup_{\mathbf{a}\in\mathcal{A}}[\mathbf{a},\mathbf{r}]\Big),
\end{equation}
where \([\mathbf{a},\mathbf{r}]\) is the axis-aligned box between \(\mathbf{a}\) and \(\mathbf{r}\). 
For each benchmark, \(\mathcal{R}\) denotes the empirical reference archive used for metric computation (Sec.~\ref{subsec:experimental_setup}).
Given \(\mathcal{R}\), the HV-Error of \(\mathcal{A}\) is:
\begin{equation}
\label{eq:hv-error}
HV\text{-}Error(\mathcal{A},\mathcal{R})=
\frac{HV(\mathcal{R};\mathbf{r})-HV(\mathcal{A};\mathbf{r})}{HV(\mathcal{R};\mathbf{r})},
\end{equation}
and ADRS is:
\begin{equation}
\label{eq:adrs}
ADRS(\mathcal{A},\mathcal{R})=
\frac{1}{|\mathcal{R}|}\sum_{\mathbf{y}\in\mathcal{R}}\min_{\mathbf{a}\in\mathcal{A}} d(\mathbf{y},\mathbf{a}),
\end{equation}
where $d(\mathbf{y},\mathbf{a})=\max_{\ell}\max\!\big(0,(a_{\ell}-y_{\ell})/y_{\ell}\big)$ is the one-sided relative Chebyshev distance over the objectives, so points that match or dominate a reference point contribute zero.
We also report feasible rate, the fraction of evaluated configurations that are clean-feasible, as a measure of sampling efficiency.

\subsection{Phase-Structured Search and MCTS}
\label{subsec:mcts}
We formulate physical design tuning as a search over configuration prefixes rather than complete solutions, since parameters are fixed stage by stage and later outcomes depend on earlier decisions~\cite{ajayi2019openroad,agnesina2023parameter}. 
The prefix state up to phase \(p\) is:
\begin{equation}
\label{eq:prefix-configuration}
\mathbf{z}^{(p)}=\big(\mathbf{x}^{(1)},\dots,\mathbf{x}^{(p)}\big).
\end{equation}
Under this view, a root-to-leaf path corresponds to a complete multi-phase configuration.

MCTS is a natural backbone for search over such partial decision states~\cite{browne2012mcts}.
Although the number of phases is modest, each phase exposes a conditional action set whose product space far exceeds any realistic evaluation budget, and the utility of a choice is revealed only after downstream stages complete.
The difficulty is therefore not search depth but allocating a limited budget across many locally plausible next-step decisions.
A tree over phase prefixes addresses this by branching only on the next unresolved phase.
It never expands all remaining phases at once, yet each evaluation still covers the full flow.
Standard MCTS iterates selection, expansion, simulation, and backpropagation.
A common selection rule, UCT~\cite{kocsis2006uct}, adds to the average branch reward an exploration bonus that grows for less-visited branches.
In our setting, each measured full-flow outcome is backpropagated along the visited path as its contribution to the trade-off frontier.

\subsection{LLMs as Guidance for Optimization}
\label{subsec:llm_guidance}

LLMs can be used as auxiliary guidance modules inside an optimization loop rather than as standalone search procedures~\cite{yang2024large}.
Conditioned on the objective, parameter semantics, and search history, they can propose candidates informed by the observed search state.
This is valuable in expensive optimization, where uninformed exploration is costly~\cite{guo2025llm}.
A complementary line of work instead retrieves tuning knowledge from a corpus of previously optimized designs~\cite{pan2025crop}.
Such retrieval requires cross-design tuning data, whereas we tune a single design from scratch without such a corpus.
LLMs are especially useful in physical design tuning, where promising parameter ranges vary across phases and partial design states, making hand-designed proposal rules difficult to generalize.

\begin{figure*}[t]
\begin{center}
\includegraphics[width=165mm]{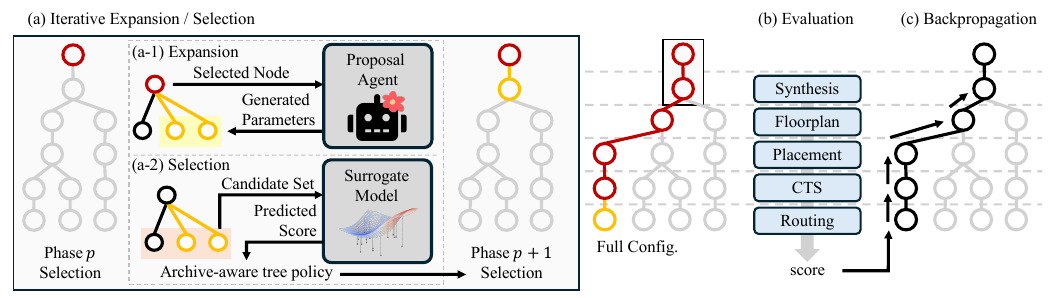}
\vspace{-0.14cm}
\caption{Overview of PILOT-Agent. (a)~The Proposal Agent expands the selected prefix, and the surrogate-scored candidates are selected through the archive-aware tree policy (Eqs.~(\ref{eq:tree_acquisition_compact})--(\ref{eq:tree_puct_compact})). (b)~The completed configuration is executed through the flow. (c)~The measured outcome is backpropagated along the visited path.}
\label{fig:pipeline-overview}
\end{center}
\vspace{-0.2cm}
\end{figure*}

\subsection{Surrogate-Assisted Screening under Expensive Evaluation}
\label{subsec:surrogate_background}

When each evaluation is expensive, surrogate models are widely used to predict the outcomes of unevaluated candidates.
Bayesian optimization embeds them in acquisition functions that trade predicted quality against uncertainty~\cite{snoek2012practical,geng2023ptpt}, and surrogate-assisted evolutionary methods use them to pre-select which offspring deserve real evaluations~\cite{jin2011surrogate}.
Physical design tuning shapes this role in two ways.
First, a configuration may fail only late in the flow, so the surrogate must predict clean-feasibility alongside quality.
Second, the target is a Pareto archive rather than a scalar optimum, so the relevant screening signal is the expected hypervolume gain of a candidate rather than predicted objective values alone~\cite{emmerich2006single}.

\section{Proposed Method} \label{sec3}

We present PILOT-Agent, a phase-aware framework for multi-objective physical design tuning.
In a conventional LLM tuning agent, the LLM both proposes the next configuration and decides that it will be executed.
PILOT-Agent separates these two roles.
The LLM only proposes candidates, and the archive-aware tree policy decides which candidate is executed.
This policy ranks each candidate by its predicted contribution to the clean-feasible Pareto archive and by the measured outcomes of its branch.
The predictions come from a lightweight surrogate refit on the runs of the current design only, so no prior training data is required.
Fig.~\ref{fig:pipeline-overview} and Alg.~\ref{alg:pilot-agent} summarize the resulting optimization loop.
Starting from a measured baseline run, PILOT-Agent initializes the phase-prefix tree~$\mathcal{T}$, surrogate model~$\mathcal{M}$, and clean-feasible archive~$\mathcal{A}$.
Each round fills a batch with LLM-proposed, surrogate-screened candidates.
The batch is then executed through the flow, and $\mathcal{T}$, $\mathcal{M}$, and $\mathcal{A}$ are updated from the measured outcomes.

\begin{algorithm}[t]
\caption{PILOT-Agent Optimization Loop}
\label{alg:pilot-agent}
\KwIn{baseline config. $\mathbf{x}_{\mathrm{base}}$, run budget $B$, batch size $b$}
\KwOut{measured clean-feasible archive $\mathcal{A}$}
$(\mathcal{T},\, \mathcal{M},\, \mathcal{A}) \gets \textsc{Initialize}(\mathbf{x}_{\mathrm{base}})$\;
$n \gets 0$\;

\While{$n < B$}{
    $\mathcal{Q} \gets \emptyset$\;
    \For{$j = 1$ \KwTo $\min(b,\; B - n)$}{
        $\mathbf{x}_j \gets \textsc{SelectRollout}(\mathcal{T},\; \mathcal{M},\; \widetilde{\mathcal{A}}(\mathcal{Q}))$\tcp*{Alg.~\ref{alg:select-expand}}
        $\mathcal{Q} \gets \mathcal{Q} \cup \{\mathbf{x}_j\}$\;
    }
    $\mathcal{Y} \gets \textsc{ExecuteBatch}(\mathcal{Q})$\;
    $\mathcal{A} \gets \textsc{UpdateArchive}(\mathcal{A},\; \mathcal{Y})$\;
    $\mathcal{M} \gets \textsc{UpdateSurrogate}(\mathcal{M},\; \mathcal{Y})$\;
    $\mathcal{T} \gets \textsc{Backpropagate}(\mathcal{T},\; \mathcal{Y})$\;
    $n \gets n + |\mathcal{Q}|$\;
}
\Return $\mathcal{A}$\;
\end{algorithm}

\subsection{Phase-Prefix Search and Candidate Proposal}
\label{subsec:llm_guided_proposal}

The search state is the phase prefix $\mathbf{z}^{(p)}$ of Sec.~\ref{subsec:mcts}, extended by one phase per tree level.
At each expandable prefix node \(u\), the \emph{Proposal Agent} suggests next-phase assignments represented as candidate children \(c\in\mathcal{C}(u)\), which are merged with the existing children.
Each child of a prefix node is a concrete assignment of all parameters of the next phase, not a discretized value of a single parameter.
Continuous ranges are therefore explored by sampling values on demand rather than by enumerating a grid.
A node requests new proposals only when its visit count is large relative to its number of children and otherwise selects among its existing children, so its child set grows as the node is revisited.
Because a tree level corresponds to a phase rather than a parameter, adding parameters enlarges each proposal but not the depth of the tree.

Measured outcomes are summarized as text in the prompt after each batch, so the agent receives feedback in context (Sec.~\ref{subsec:llm_guidance}) and the LLM itself is never retrained.
The prompt contains a generic description and tuning hint for each parameter, the objective, the current prefix, and summaries of measured runs, including frontier-improving runs, failure-prone regions, under-explored trade-off regions, and near-duplicate configurations.
The goal is not a single best continuation but a compact candidate set that balances refinement near good regions with exploration of under-represented directions.
For each \(c\), \(\mathbf{x}_c\) denotes the full configuration obtained by completing the remaining phases through that child.
We call \(\mathbf{x}_c\) a \emph{rollout}, which is scored by the surrogate and executed only if selected for a batch slot.
Line~8 of Alg.~\ref{alg:select-expand} scores each rollout with the acquisition $\Phi$, value $V_u$, and traversal score $S_u$ of Sec.~\ref{subsec:archive_aware_tree_policy}.

\begin{algorithm}[t]
\caption{\textsc{SelectRollout}: Archive-Aware Traversal}
\label{alg:select-expand}
\KwIn{tree $\mathcal{T}$, surrogate $\mathcal{M}$, valuation archive $\widetilde{\mathcal{A}}$}
\KwOut{selected candidate configuration}

$u \gets \textsc{Root}(\mathcal{T})$\;

\While{$u$ is expandable}{
    $\Gamma_u \gets \textsc{BuildLocalContext}(u,\widetilde{\mathcal{A}})$\;
    $\mathcal{C}_{\mathrm{new}}(u) \gets \textsc{ProposalAgent}(u,\Gamma_u)$\;
    $\mathcal{C}(u) \gets \textsc{Children}(u)\cup\mathcal{C}_{\mathrm{new}}(u)$\;

    \ForEach{$c\in\mathcal{C}(u)$}{
        $\mathbf{x}_c \gets \textsc{CompleteRollout}(u,c)$\;
        compute $\Phi(c)$, $V_u(c)$, $S_u(c)$\tcp*{Eqs.~(\ref{eq:tree_acquisition_compact})--(\ref{eq:tree_puct_compact})}
    }

    $c^\star \gets \arg\max_{c\in\mathcal{C}(u)} S_u(c)$\;

    \uIf{$c^\star \in \textsc{Children}(u)$}{
        $u \gets c^\star$\;
    }
    \Else{
        \Return $\mathbf{x}_{c^\star}$\;
    }
}
\Return{} the completed configuration of $u$\;
\end{algorithm}

\subsection{Archive-Aware Tree Policy}
\label{subsec:archive_aware_tree_policy}

\begin{figure}[t]
\begin{center}
\vspace{-0.25cm}
\includegraphics[width=78mm]{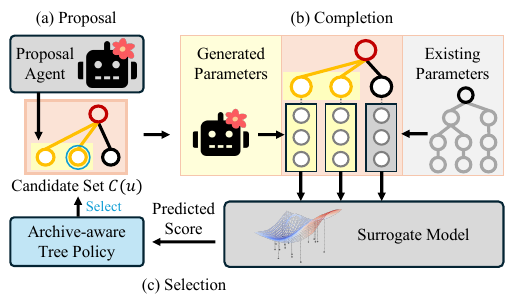}
\vspace{-0.5cm}
\caption{The \textsc{SelectRollout} procedure (Alg.~\ref{alg:select-expand}). (a)~The Proposal Agent proposes next-phase candidates. (b)~Each candidate is completed with generated and existing parameters. (c)~The surrogate scores the rollouts and the tree policy selects one.}
\label{fig:selectandexpand}
\end{center}
\vspace{-0.5cm}
\end{figure}

The archive-aware tree policy scores each candidate in~$\mathcal{C}(u)$ through Eqs.~(\ref{eq:tree_acquisition_compact})--(\ref{eq:tree_puct_compact}) before branch selection (Fig.~\ref{fig:selectandexpand}).
The acquisition estimates how much executing $\mathbf{x}_c$ would improve the archive.
A failed run adds nothing and a clean-feasible run adds its hypervolume gain, so the expected improvement is the feasibility probability times the expected gain over the archive $\widetilde{\mathcal{A}}$:
\begin{equation}
\label{eq:tree_acquisition_compact}
\Phi(c) =
p_{\mathrm{feas}}(\mathbf{x}_c)\,
G(\mathbf{x}_c;\widetilde{\mathcal{A}})
+\lambda_{\mathrm{cov}}\,U(\mathbf{x}_c)
-\lambda_{\mathrm{dup}}\,D(\mathbf{x}_c).
\end{equation}
This product is the feasibility-weighted expected hypervolume improvement used in constrained multi-objective Bayesian optimization~\cite{gardner2014constrained,emmerich2006single}.
The terms \(U(\mathbf{x}_c)\) and \(D(\mathbf{x}_c)\) reward under-covered trade-off regions and penalize redundant rollouts, respectively, with nonnegative weights \(\lambda_{\mathrm{cov}}\) and \(\lambda_{\mathrm{dup}}\).
Because the expected gain alone favors regions near already strong points, $U$ and $D$ spread evaluations along the frontier and avoid near-duplicate rollouts.

The exploitation value blends this acquisition estimate with branch-level evidence accumulated from measured executions:
\begin{equation}
\label{eq:tree_value_compact}
V_u(c)=
\begin{cases}
\Phi(c), & c\in\mathcal{C}_{\mathrm{new}}(u),\\[3pt]
(1-\beta_{u,c})H_{u,c}+\beta_{u,c}\Phi(c),
& c\in\textsc{Children}(u).
\end{cases}
\end{equation}
where \(H_{u,c}\) is an exponentially weighted average of the realized hypervolume gains from measured rollouts through branch \(c\), and \(\beta_{u,c}\in(0,1]\) decreases with the number of measured rollouts through the branch.
$\Phi(c)$ predicts and $H_{u,c}$ measures the same quantity, the hypervolume gain of a rollout through $c$.
Blending them lets a branch rely on the surrogate while it has few measurements and increasingly on its own measured outcomes afterwards.
Because $G$ integrates over the surrogate posterior, candidates with uncertain predictions are not dismissed early, and an optimistic misprediction costs executions only until the branch accumulates measured history.
Eqs.~(\ref{eq:tree_acquisition_compact}) and~(\ref{eq:tree_value_compact}) together constitute the \emph{archive-aware value} of the tree policy.

The final traversal score augments the exploitation value with visit-based exploration:
\begin{equation}
\label{eq:tree_puct_compact}
S_u(c) = V_u(c) + \lambda_{\mathrm{occ}}\,\Omega_u(c) + \alpha\,\frac{\sqrt{N_u}}{1+N_{u,c}},
\end{equation}
where \(\Omega_u(c)\) is an occupancy regularizer that favors less-visited branches, \(N_u\) and \(N_{u,c}\) are the node and branch visit counts, and \(\lambda_{\mathrm{occ}}\) and \(\alpha\) weight the last two terms.

\begin{table}[b]
\centering
\vspace{-0.2cm}
\caption{Tunable parameters organized by design phase.}
\label{tab:parameters}
\vspace{-0.3cm}
\renewcommand{\tabcolsep}{0.35mm}
{\small
\begin{tabular}{clll}
\toprule
 & Parameter & Description & Range / Type \\
\hline
\cmidrule(){1-4}
\multirow{2}{*}{1$\;$}  & Clock Period     & Target clock period                        & $[T_l, T_u]$ / F \\
                        & Flatten Hierarchy & Enable hierarchical synthesis             & $\{0, 1\}$ / B \\
\cmidrule(){1-4}
\multirow{3}{*}{2$\;$}  & Core Utilization & Target density of cells in core area       & $[20, 99]$ / I \\
                        & Aspect Ratio & Height-to-width ratio of core area             & $[0.1, 2.0]$ / F \\
                        & Density Margin & Lower bound adjustment for density           & $[0.00, 0.99]$ / F \\
\cmidrule(){1-4}
\multirow{2}{*}{3$\;$}  & Global Padding     & Cell padding for global placement        & $[0, 4]$ / I \\
                        & Detailed Padding   & Cell padding for detailed placement      & $[0, 4]$ / I \\
\cmidrule(){1-4}
\multirow{2}{*}{4$\;$}  & Cluster Size      & Maximum sinks per CTS cluster             & $[10, 40]$ / I \\
                        & Cluster Diameter   & Maximum diameter of CTS cluster          & $[50, 200]$ / I \\
\cmidrule(){1-4}
\multirow{2}{*}{5$\;$}  & Layer Adjust       & Factor for routing layer capacity        & $[0.00, 0.99]$ / F \\
                        & TNS End Percent    & Percentage of timing violations to fix   & $[0, 100]$ / I \\
\bottomrule
\multicolumn{4}{l}{\multirow{1.2}{*}{\footnotesize 1: Synthesis / 2: Floorplan / 3: Placement / 4: CTS / 5: Routing}} \\
\multicolumn{4}{l}{\multirow{1.2}{*}{\footnotesize F: Float / I: Integer / B: Binary $\quad T_l\,/\,T_u$: Lower / upper bound of clock period}} \\
\end{tabular}
}
\vspace{-0.15cm}
\end{table}

\subsection{Surrogate Screening and Batch Construction}
\label{subsec:triage}

Each batch slot takes the rollout that maximizes $S_u(c)$ in Sec.~\ref{subsec:archive_aware_tree_policy}.
Duplicates of executed or already reserved configurations are filtered before reservation, and the budget \(B\) counts executed configurations only.
Batches are built sequentially rather than by picking independent top-ranked candidates.
At each batch slot, the provisional archive $\widetilde{\mathcal{A}}(\mathcal{Q})$ augments the measured archive $\mathcal{A}$ with the objective vectors $\hat{\mathbf{f}}(\mathbf{x})$ that the surrogate predicts for rollouts already reserved in $\mathcal{Q}$.
This discourages redundant reservations and promotes complementary trade-off directions.
This greedy construction mirrors sequential batch expected-hypervolume optimization~\cite{daulton2020qehvi}.
The provisional archive affects only within-batch proposal and selection.
The measured archive, surrogate, and tree statistics are updated only after batch execution.

\begin{table*}[t]
\begin{center}
\caption{Clean-feasible frontier recovery across benchmark--technology pairs. Lower HV-Error and ADRS are better, while higher feasible rate is better. $\pm$ denotes the standard deviation, and the best HV-Error and ADRS per benchmark are in bold.}
\label{tab:main_results}
\vspace{-0.25cm}
\renewcommand{\tabcolsep}{0.85mm}
{\small
\begin{tabular}{cccccccccccccc}
\toprule
\multirow{2.5}{*}{Platform} & \multirow{2.5}{*}{Design} & \multicolumn{3}{c}{AutoTuner\cite{jung2021metrics}} & \multicolumn{3}{c}{REMOTune\cite{zheng2023boosting}} & \multicolumn{3}{c}{ORFS-Agent\cite{ghose2025orfsagent}} & \multicolumn{3}{c}{PILOT-Agent (Ours)} \\
\cmidrule(l){3-5}\cmidrule(l){6-8}\cmidrule(l){9-11}\cmidrule(l){12-14}
 &  & HV-Err. & ADRS & Feas.~(\%) & HV-Err. & ADRS & Feas.~(\%) & HV-Err. & ADRS & Feas.~(\%) & HV-Err. & ADRS & Feas.~(\%) \\
\hline
\cmidrule(){1-14}
\multirow{3}{*}{SKY130HD}
 & AES  & 0.205{\footnotesize$\pm$.023} & 0.099{\footnotesize$\pm$.048} & 21.1 & 0.191{\footnotesize$\pm$.028} & 0.074{\footnotesize$\pm$.019} & 26.6 & 0.109{\footnotesize$\pm$.031} & 0.043{\footnotesize$\pm$.013} & 84.1 & \textbf{0.089}{\footnotesize$\pm$.010} & \textbf{0.032}{\footnotesize$\pm$.002} & 96.1 \\
 & JPEG & 0.348{\footnotesize$\pm$.052} & 0.077{\footnotesize$\pm$.010} & 38.5 & 0.148{\footnotesize$\pm$.045} & 0.028{\footnotesize$\pm$.009} & 68.5 & 0.345{\footnotesize$\pm$.140} & 0.123{\footnotesize$\pm$.043} & 84.6 & \textbf{0.115}{\footnotesize$\pm$.001} & \textbf{0.021}{\footnotesize$\pm$.002} & 81.5 \\
 & IBEX & 0.124{\footnotesize$\pm$.025} & 0.039{\footnotesize$\pm$.004} & 32.8 & 0.288{\footnotesize$\pm$.201} & 0.064{\footnotesize$\pm$.032} & 41.9 & 0.183{\footnotesize$\pm$.037} & 0.133{\footnotesize$\pm$.031} & 80.5 & \textbf{0.061}{\footnotesize$\pm$.016} & \textbf{0.019}{\footnotesize$\pm$.002} & 79.4 \\
\cmidrule(l){1-14}
\multirow{3}{*}{ASAP7}
 & AES  & 0.151{\footnotesize$\pm$.018} & 0.032{\footnotesize$\pm$.003} & 45.1 & 0.115{\footnotesize$\pm$.001} & 0.021{\footnotesize$\pm$.001} & 78.1 & 0.132{\footnotesize$\pm$.025} & 0.046{\footnotesize$\pm$.005} & 98.7 & \textbf{0.069}{\footnotesize$\pm$.008} & \textbf{0.012}{\footnotesize$\pm$.002} & 98.2 \\
 & JPEG & 0.287{\footnotesize$\pm$.033} & 0.083{\footnotesize$\pm$.004} & 37.8 & \textbf{0.157}{\footnotesize$\pm$.016} & \textbf{0.018}{\footnotesize$\pm$.001} & 73.4 & 0.163{\footnotesize$\pm$.033} & 0.070{\footnotesize$\pm$.017} & 88.0 & 0.311{\footnotesize$\pm$.075} & 0.031{\footnotesize$\pm$.006} & 94.3 \\
 & IBEX & 0.247{\footnotesize$\pm$.036} & 0.613{\footnotesize$\pm$.402} & 22.9 & 0.251{\footnotesize$\pm$.078} & 0.072{\footnotesize$\pm$.056} & 35.7 & 0.122{\footnotesize$\pm$.038} & 0.620{\footnotesize$\pm$.413} & 47.7 & \textbf{0.087}{\footnotesize$\pm$.001} & \textbf{0.018}{\footnotesize$\pm$.002} & 61.2 \\
\cmidrule(){1-14}
\multicolumn{2}{c}{Average} & 0.227 & 0.157 & 33.0 & 0.191 & 0.046 & 54.0 & 0.176 & 0.172 & 80.6 & \textbf{0.122} & \textbf{0.022} & 85.1 \\
\bottomrule
\end{tabular}
}
\end{center}
\vspace{-0.2cm}
\end{table*}

\section{Experimental Results} \label{sec4}

\subsection{Experimental Setup}
\label{subsec:experimental_setup}

We tune 11 parameters (Table~\ref{tab:parameters}) spanning synthesis, floorplanning, placement, CTS, and routing on OpenROAD-flow-scripts (ORFS)~\cite{ajayi2019openroad} release 26Q1, across AES, JPEG, and IBEX on SKY130HD~\cite{sky130hd} and ASAP7~\cite{vashishtha2017asap7}.
Parameter ranges follow each benchmark's ORFS baseline and cover valid per-phase operating regions.
All methods run under the same budget of $B=128$ full-flow executions with batch size $4$, the same per-benchmark baseline, and the same termination rules, over three seeds, with timeout, crash, and invalid-flow outcomes counted as failed runs.
Each run is allocated four CPU cores and 4\,GB of RAM on an Intel Xeon Gold 5218.

For PILOT-Agent, the Proposal Agent uses GPT-5.4-nano (zero-shot, temperature $T=0.55$, up to 6 candidates per call), and backbone sensitivity is analyzed in Sec.~\ref{subsec:ablation}.
The surrogate uses three Gaussian-process regressors for the posterior over power, ECP, and area and random-forest classifiers for clean-feasibility.
The gain term $G(\mathbf{x}_c;\widetilde{\mathcal{A}})$ is the expected nonnegative hypervolume improvement estimated from posterior samples of the objective model.
We compare against AutoTuner~\cite{jung2021metrics} (derivative-free multi-fidelity BO in ORFS), REMOTune~\cite{zheng2023boosting} (random-embedding multi-objective trust-region BO), and ORFS-Agent~\cite{ghose2025orfsagent} (LLM-agent tuning on the same GPT-5.4-nano backbone), spanning the three dominant flow-tuning paradigms.
All baselines run through a shared harness that reuses our parameter space, ORFS execution path, and metric computation.
ORFS-Agent follows its published workflow and is prompted only with the search space, baseline, objective, and run history, whereas our Proposal Agent also receives the generic parameter hints of Sec.~\ref{subsec:llm_guided_proposal}. 
The ablation in Sec.~\ref{subsec:ablation} separates the effect of search structure from this difference.
For each benchmark, the reference archive $\mathcal{R}$ used in Eqs. (\ref{eq:hv-error}) and (\ref{eq:adrs}) is the nondominated set pooled over the clean-feasible solutions of all compared methods and seeds. 
The same $\mathcal{R}$ and reference point score every method and ablation.
Prompts, surrogate settings, baseline adaptations, and hyperparameters are in the released code.

\begin{figure}[t]
\begin{center}
\vspace{-0.15cm}
\includegraphics[width=83mm]{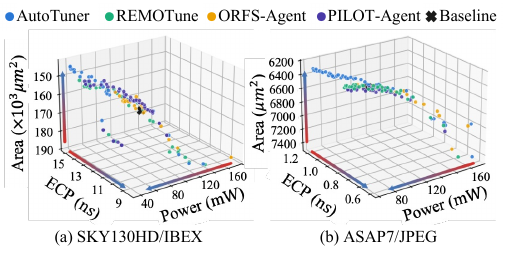}
\vspace{-0.3cm}
\caption{Nondominated clean-feasible points recovered on (a) SKY130HD/IBEX and (b) ASAP7/JPEG. Arrows indicate the improving direction of each objective.}
\label{fig:pareto-result}
\end{center}
\vspace{-0.2cm}
\end{figure}

\begin{table*}[t]
\begin{center}
\caption{Best clean-feasible value of each objective when the other two objectives are constrained to be no worse than the default OpenROAD baseline (Base.). `---': no clean-feasible point satisfies the constraints.}
\label{tab:constrained_ppa}
\vspace{-0.25cm}
\renewcommand{\tabcolsep}{1.63mm}
{\small
\begin{tabular}{ccccccccccccccccc}
\toprule
\multirow{2.5}{*}{Platform} & \multirow{2.5}{*}{Design} & \multicolumn{5}{c}{Best Power (mW)} & \multicolumn{5}{c}{Best ECP (ns)} & \multicolumn{5}{c}{Best Area ($\mu m^2$)} \\
\cmidrule(l){3-7}\cmidrule(l){8-12}\cmidrule(l){13-17}
 &  & Base. & \cite{jung2021metrics} & \cite{zheng2023boosting} & \cite{ghose2025orfsagent} & Ours & Base. & \cite{jung2021metrics} & \cite{zheng2023boosting} & \cite{ghose2025orfsagent} & Ours & Base. & \cite{jung2021metrics} & \cite{zheng2023boosting} & \cite{ghose2025orfsagent} & Ours \\
\hline
\cmidrule(){1-17}
\multirow{3}{*}{SKY130HD}
 & AES  & 405.9 & 1197.6 & 426.5 & 396.8 & 398.2 & 4.400 & 4.460 & 4.798 & 4.346 & 4.298 & 111716 & --- & --- & 108125 & 107597 \\
 & JPEG & 515.4 & --- & --- & 456.7 & 466.5 & 5.490 & 5.834 & 5.501 & 5.364 & 5.369 & 470311 & 523689 & 505514 & 443609 & 447479 \\
 & IBEX & 70.0 & 62.3 & 63.3 & 64.2 & 63.0 & 10.130 & 10.073 & 9.851 & 9.822 & 9.686 & 157149 & 151268 & 152155 & 150721 & 150977 \\
\cmidrule(l){1-17}
\multirow{3}{*}{ASAP7}
 & AES  & 144.8 & --- & 139.4 & 130.3 & 128.7 & 0.424 & 0.435 & 0.422 & 0.410 & 0.405 & 1927.8 & 2047.8 & 1879.6 & 1840.7 & 1815.1 \\
 & JPEG & 88.6 & 94.2 & 67.0 & 79.1 & 66.6 & 0.853 & 0.877 & 0.803 & 0.803 & 0.788 & 6359.2 & --- & 6275.7 & 6311.5 & 6315.9 \\
 & IBEX & 28.4 & --- & --- & 26.0 & 27.7 & 1.082 & 1.102 & 1.121 & 1.075 & 1.073 & 2490.2 & --- & --- & 2447.0 & 2471.4 \\
\bottomrule
\end{tabular}
}
\end{center}
\vspace{-0.2cm}
\end{table*}

\subsection{Main Results}
\label{subsec:main_results}

\subsubsection{Pareto Frontier Recovery}
\label{subsec:ppa_results}
Table~\ref{tab:main_results} shows that PILOT-Agent achieves the lowest HV-Error and ADRS on five of the six pairs.
Its mean HV-Error of 0.122 is 31\% below the best baseline average (0.176, ORFS-Agent), and its mean ADRS of 0.022 is 52\% below that of REMOTune (0.046).
It also attains the highest average feasible rate (85.1\% vs.\ 80.6\%) and the lowest HV-Error standard deviation on four of the six pairs.
Fig.~\ref{fig:pareto-result} illustrates both a representative gain and the sole exception.
On SKY130HD/IBEX, PILOT-Agent nearly halves the best baseline HV-Error (0.061 vs.\ 0.124) while recovering a broader clean-feasible frontier.
On ASAP7/JPEG, REMOTune achieves lower HV-Error (0.157 vs.\ 0.311), whereas PILOT-Agent maintains substantially higher feasibility (94.3\% vs. 73.4\%) with ADRS values of 0.031 and 0.018, respectively.
This gap is concentrated in a few low-ECP extremes that consume more than $1.4\times$ the baseline power, rather than spread across the frontier.

As a robustness check, we recompute all methods' metrics against a common baseline-only reference archive that excludes PILOT-Agent while keeping the hypervolume reference point fixed.
PILOT-Agent remains best on the same five benchmarks, with average HV-Error 0.121 and ADRS 0.027, indicating that the ranking is not driven by its contribution to the pooled reference archive.

\begin{figure}[t]
\begin{center}
\includegraphics[width=85mm]{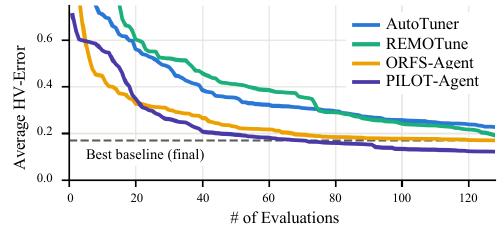}
\vspace{-0.7cm}
\caption{Benchmark-averaged HV-Error versus executed full-flow evaluations. The dashed line marks the final HV-Error of the best baseline.}
\label{fig:convergence}
\end{center}
\vspace{-0.3cm}
\end{figure}

\begin{figure}[t]
\begin{center}
\vspace{-0.35cm}
\includegraphics[width=85mm]{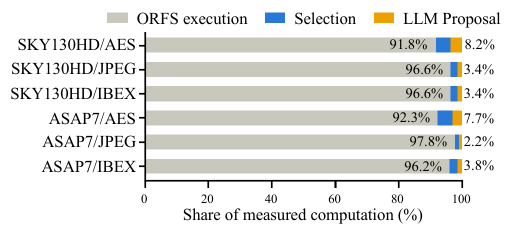}
\vspace{-0.7cm}
\caption{Runtime composition of PILOT-Agent per benchmark, normalized to the measured computation.}
\label{fig:cost}
\end{center}
\vspace{-0.1cm}
\end{figure}

\subsubsection{Practical PPA under Constraints}
\label{subsec:constrained_ppa}

To check whether the recovered solutions are useful design points, Table~\ref{tab:constrained_ppa} reports the best clean-feasible value of each objective when the other two are no worse than the default OpenROAD baseline.
PILOT-Agent attains the best value in nine of 18 cases, stays within 1\% of the best in fifteen, and always yields a qualifying point, whereas AutoTuner and REMOTune fail to qualify in six and four cases.
ORFS-Agent, the strongest baseline, is best in seven cases, but its lead exceeds 2\% only twice.
On ASAP7/JPEG, the one frontier-metric loss, PILOT-Agent attains the best constrained power and ECP and stays within 1\% on area, so its HV-Error deficit lies outside the constrained region.

\subsubsection{Convergence and Optimization Cost}
\label{subsec:convergence}
\label{subsec:cost}

Since the motivation is expensive evaluation, final-budget snapshots alone do not establish budget efficiency.
Fig.~\ref{fig:convergence} plots benchmark-averaged best-so-far HV-Error against the number of executed full-flow evaluations.
On the five pairs where PILOT-Agent attains the lowest final HV-Error, it reaches the final HV-Error of the best baseline after 60 evaluations on average, or 47\% of the budget.

Fig.~\ref{fig:cost} reports the runtime composition of PILOT-Agent per benchmark, normalized to the measured computation.
ORFS execution accounts for 96.0\% of the measured runtime overall, candidate selection for 2.3\%, and LLM proposal generation for 1.6\%. 
Averaged over benchmark--seed optimization campaigns, each 128-evaluation campaign uses approximately 150 LLM calls and 1.0 million total input-plus-output tokens for proposal generation. 
Thus, EDA execution remains the dominant cost, and the combined method-side share does not exceed 8.2\% on any benchmark.

\begin{table}[t]
\begin{center}
\caption{Component ablation under the same budget, scored against the reference archives of Table~\ref{tab:main_results}. $\pm$ is the standard deviation.}
\label{table:ablation}
\vspace{-0.3cm}
\renewcommand{\tabcolsep}{0.56mm}
{\small
\begin{tabular}{lcccccc}
\toprule
\multirow{2.5}{*}{Method} & \multicolumn{3}{c}{SKY130HD/IBEX} & \multicolumn{3}{c}{ASAP7/AES} \\
\cmidrule(l){2-4}\cmidrule(l){5-7}
 & HV-Err. & ADRS & Feas. & HV-Err. & ADRS & Feas. \\
\hline
\cmidrule(){1-7}
\multicolumn{7}{l}{\emph{LLM proposal}} \\
\hspace{0.8em}w/o LLM       & 0.141{\footnotesize$\pm$.004} & 0.111{\footnotesize$\pm$.000} & 62.2 & 0.221{\footnotesize$\pm$.005} & 0.051{\footnotesize$\pm$.001} & 89.3 \\
\hspace{0.8em}w/o Context   & 0.069{\footnotesize$\pm$.012} & 0.024{\footnotesize$\pm$.003} & 69.0 & 0.069{\footnotesize$\pm$.007} & 0.013{\footnotesize$\pm$.001} & 98.4 \\
\multicolumn{7}{l}{\emph{Surrogate screening}} \\
\hspace{0.8em}w/o Surrogate & 0.106{\footnotesize$\pm$.018} & 0.034{\footnotesize$\pm$.004} & 68.5 & 0.074{\footnotesize$\pm$.007} & 0.013{\footnotesize$\pm$.001} & 98.2 \\
\multicolumn{7}{l}{\emph{Search structure}} \\
\hspace{0.8em}Flat LLM      & 0.098{\footnotesize$\pm$.014} & 0.029{\footnotesize$\pm$.008} & 85.2 & 0.082{\footnotesize$\pm$.015} & 0.016{\footnotesize$\pm$.003} & 99.0 \\
\cmidrule(){1-7}
Full (\footnotesize PILOT-Agent) & 0.061{\footnotesize$\pm$.016} & 0.019{\footnotesize$\pm$.002} & 79.4 & 0.069{\footnotesize$\pm$.008} & 0.012{\footnotesize$\pm$.002} & 98.2 \\
\bottomrule
\end{tabular}
}
\end{center}
\vspace{0.1cm}
\end{table}

\subsection{Ablation Study}
\label{subsec:ablation}

\textbf{Component ablation.}
Table~\ref{table:ablation} evaluates four ablated variants under the same budget and reference archives on two contrasting benchmarks: SKY130HD/IBEX, where clean-feasibility is relatively scarce, and ASAP7/AES, where it is near-saturated.
\emph{w/o LLM} replaces the Proposal Agent with random proposals over the same parameter ranges; \emph{w/o Context} removes measured search-state feedback from the prompt; \emph{w/o Surrogate} executes proposals in generation order after deduplication; and \emph{Flat LLM} removes the phase-prefix tree and proposes complete configurations.
Full reproduces the corresponding entries of Table~\ref{tab:main_results}.
Removing informed LLM proposals causes the largest degradation in both regimes, raising HV-Error from 0.061 to 0.141 on SKY130HD/IBEX and from 0.069 to 0.221 on ASAP7/AES.
Because this loss persists even when feasibility is near-saturated, informed proposals improve frontier quality rather than merely the number of successful runs.
The effects of context and surrogate screening depend on the feasibility regime.
On SKY130HD/IBEX, w/o Context and w/o Surrogate reach similar feasible rates (69.0\% and 68.5\%), yet w/o Surrogate has much worse HV-Error (0.106 vs.\ 0.069), indicating that screening decides which candidates consume the budget rather than only filtering likely failures.
On ASAP7/AES, both variants remain close to Full.

Flat LLM uses the same parameter hints and surrogate acquisition as Full but proposes complete configurations without the phase-prefix tree.
It attains the highest feasible rate on SKY130HD/IBEX (85.2\%) but degrades HV-Error and ADRS from 0.061 and 0.019 to 0.098 and 0.029, and it also trails Full on ASAP7/AES.
Because the parameter hints and surrogate are held fixed, this gap isolates the contribution of phase-prefix search, showing that proposing complete configurations favors safe evaluations but covers the frontier less well.
The margin over ORFS-Agent in Table~\ref{tab:main_results}, by contrast, reflects both search structure and prompt content.
Together, these results indicate complementary roles: the LLM broadens the candidate set, the surrogate screens candidates before execution, and the phase-prefix tree organizes frontier-oriented exploration.

\textbf{Backbone sensitivity.}
Table~\ref{table:backbone} replaces GPT-5.4-nano with DeepSeek-V4-flash on SKY130HD/IBEX.
HV-Error rises from 0.061 to 0.087, still below the best baseline (0.124) and well below random proposals (0.141).
PILOT-Agent thus depends more on informed proposals and search organization than on the specific backbone.

\begin{table}[t]
\begin{center}
\caption{Backbone sensitivity of PILOT-Agent on SKY130HD/IBEX. $\pm$ denotes the standard deviation.}
\label{table:backbone}
\vspace{-0.3cm}
\renewcommand{\tabcolsep}{3.5mm}
{\small
\begin{tabular}{lccc}
\toprule
Method & HV-Err. & ADRS & Feas. (\%) \\
\hline
\cmidrule(){1-4}
GPT-5.4-nano   & 0.061{$\pm$.016} & 0.019{$\pm$.002} & 79.4 \\
DeepSeek-V4-flash  & 0.087{$\pm$.015} & 0.026{$\pm$.007} & 71.4 \\
\bottomrule
\end{tabular}
}
\end{center}
\vspace{-0.1cm}
\end{table}

\textbf{Limitations.}
The ablation covers two benchmarks chosen as contrasting feasibility regimes, so component-wise conclusions are not verified on the other four.
We do not measure how surrogate prediction quality evolves with the budget, although the blending in Eq.~(\ref{eq:tree_value_compact}) limits the effect of early errors.
Per-phase branching keeps the tree depth fixed as parameters are added, but surrogate reliability beyond our 11-parameter space is untested.

\section{Conclusion}

PILOT-Agent organizes expensive physical design tuning around phase-prefix search.
The LLM broadens the candidate set, while surrogate estimates and measured branch outcomes guide the evaluation budget.
This organization improves clean-feasible Pareto-front recovery across diverse designs and technologies.
PILOT-Agent also reaches the best baseline frontier quality within 47\% of the evaluation budget on average.
The recovered frontiers provide practical design points for improving one PPA objective without degrading the others relative to the default baseline.

\begin{acks}
This work was supported by Samsung Electronics Co., Ltd, and by the POSTECH Institute of Artificial Intelligence.
\end{acks}

\bibliographystyle{ACM-Reference-Format}
\bibliography{reference}


\end{document}